\documentclass[aps,prc,twocolumn,showpacs,superscriptaddress]{revtex4}
\usepackage{graphics}
\usepackage{epsfig}
\usepackage[cp1251]{inputenc}
\usepackage{amssymb,latexsym,amsmath,bm}
\usepackage{color}
\usepackage[colorlinks,linkcolor=blue,citecolor=blue]{hyperref}

\newcommand{\beq}{\begin{equation}}
\newcommand{\eeq}{\end{equation}}
\newcommand{\bea}{\begin{eqnarray}}
\newcommand{\eea}{\end{eqnarray}}
\newcommand{\eps}{\varepsilon}

\newcommand{\bp}{{\bf p}}
\newcommand{\bq}{{\bf q}}
\newcommand{\br}{{\bf r}}

\newcommand{\sg}{{\bm \sigma}}

\newcommand{\pd}{\partial}

\newcommand{\dl}{\delta}
\newcommand{\Dl}{\Delta}
\newcommand{\om}{\omega}

\newcommand{\fl}{\mbox{\scriptsize FL}}

\newcommand{\eff}{\mbox{\scriptsize eff}}

\begin{document}


\title{$^1S_0$ pairing for neutrons in dense neutron matter induced by a soft pion}
\author{S.~S.~Pankratov}
\affiliation{National Research Centre Kurchatov Institute, pl. Akademika Kurchatova 1, Moscow, 123182, Russia}
\author{M.~Baldo}
\affiliation{Istituto Nazionale di Fisica Nucleare, Sezione di Catania, 64
Via S.-Sofia, I-95123 Catania, Italy}
\author{E.~E.~Saperstein}
\affiliation{National Research Centre Kurchatov Institute, pl. Akademika Kurchatova 1, Moscow, 123182, Russia}

\date{\today}

\begin{abstract}

The possibility of neutron pairing in the $^1S_0$ channel is studied for dense neutron matter in a vicinity of the $\pi^0$ condensation point. The $^1S_0$ pairing gap $\Dl$ is shown to occur in a model with a pairing force induced by the exchange of a soft neutral pionic mode. The soft pion induced potential $V_{\pi}(r)$ is characterized by an attenuating oscillatory behavior in coordinate space, while in momentum space all $S$-wave matrix elements $V_{\pi}(p,p')$ are positive. The solution of the gap equation reveals strong momentum dependence.

\end{abstract}


\maketitle

\section{Introduction}

Pairing in neutron matter in the $^1S_0$ partial wave channel is sufficiently well studied. On the level of the BCS approach with the use of a bare $NN$ potential as a pairing interaction and free single-particle energies, it is well established that the $^1S_0$ pairing correlations exist in neutron matter in a range of densities $n\le n_S\simeq0.18$ fm$^{-3}$ \cite{Dean_Hjorth-Jensen2003}. The density $n_S$ of the vanishing of the $^1S_0$ pairing corresponds to the neutron Fermi momentum $p_{\rm F}\simeq1.75$ fm$^{-1}$. Various investigations of nucleon pairing beyond BCS have shown that account of in-medium corrections to the pairing interaction and renormalization of the single-particle spectrum both lead to a rather strong reduction of the $^1S_0$ pairing gap whereas the critical value $n_S$ changes not significantly \cite{BaldoGrasso2000,LombardoSchuckZuo2001,BozekPRC2000,MutherDickhoff2005,
SchulzePollsRamos2001,ShenLombardoSchuck2005,CaoLombardoSchuck2006}. Therefore neutron superfluidity at densities $n\ge n_S$, which are relevant to a neutron star core region, $n\gtrsim 0.5\,n_0$, where $n_0\simeq0.16$ fm$^{-3}$ is the nuclear saturation density, is usually connected with the $^3P_2-^3F_2$ coupled channel \cite{Dean_Hjorth-Jensen2003,BaldoPRC1998,KhodelPRL2001}. In this work we show that the $^1S_0$ pairing is possible to reappear in dense neutron matter in a vicinity of the $\pi^0$ condensation instability \cite{Picond,MigdalRevModPhys1978,PionBook,MigdalPhysRep1990} where the collective $\pi^0$-like excitations become quit soft.

It is worth to note that the problem of a consistent description of mesonic degrees of freedom in nuclear and neutron matter in connection with the concept of pion condensation and the estimations by Migdal \cite{Picond,MigdalRevModPhys1978} was extensively discussed  in seventies and eighties \cite{discussion1,discussion2,discussion3,discussion4} with different conclusions. However, more recent microscopic considerations of neutron matter \cite{Wiringa1988,AkmalPandRav1998} gave additional arguments in favor the $\pi^0$-condensation  phenomenon in neutron stars at densities we consider. In any case, we suppose that the $\pi^0$-condensation does occur and we start our study from this assertion. The results of the analysis could reveal possible astrophysical signals sensitive to the presence of pion condensation.

In homogeneous neutron matter, the $\pi^0$ propagator is given by the familiar expression, \beq
D(\om,\bq)=\left(\om^2-q^2-m_{\pi}^2-\Pi(\om,q;n)\right)^{-1}\,,\label{Dpi0}\eeq where the density dependent polarization operator $\Pi(\om,q;n)$ accounts for in-medium pion scattering processes
including particle-hole ($N\bar N$) and $\Delta$-isobar-hole ($\Delta\bar N$) excitations. The
dispersion law $\om=\om(q)$ of the in-medium pionic field is identified with poles of the propagator (\ref{Dpi0}). It contains three branches which correspond to the $N\bar N$, $\pi^0$, and $\Delta\bar N$ degrees of freedom, which are strongly mixed. For brevity we refer to the lowest in the energy mode as a soft in-medium pion. In neutron matter the $\pi^0$ condensation occurs at the critical density $n_c$ when a zero energy pole appears in the $\pi^0$ propagator (\ref{Dpi0}) at a certain momentum $q_c$ \cite{MigdalRevModPhys1978,PionBook,MigdalPhysRep1990}. The following conditions hold at the critical point: \beq D^{-1}(0,q_c)=0\,,\quad \left.\dfrac{\pd D^{-1}(0,q)}{\pd q^2}\right|_{q_c}=0\,.\eeq We limit ourselves with density values close to the critical one. In this case, the $D$-function takes the form \cite{MigdalPhysRep1990}\beq D(\omega,\bq) \simeq -\dfrac{1}{\alpha(q^2-q_c^2)^2+\beta(n_c-n)-i\gamma |\omega|},\label{Dpi0c}\eeq as follows from the power expansion of the propagator (\ref{Dpi0}) at the critical parameters $q_c$, $n_c$ and zero energy. The coefficients $\alpha,\beta,\gamma$ are positive constants which can be calculated if a model for the pion polarization operator is suggested. The realistic estimation of the critical density for the $\pi^0$ condensation is $n_c\simeq0.2$ fm$^{-3}$, as follows from the microscopic investigations of nuclear matter \cite{Wiringa1988,AkmalPandRav1998} cited above. The critical momentum $q_c$ is not known sufficiently well and is estimated to be in the range $(0.7\div 1.0)\,p_{\rm F}$.

The presence of a soft collective mode in a Fermi system affects strongly the quasiparticle
interaction \cite{Dugaev1976}. In vicinity of the $\pi^0$ condensation point in neutron matter, the scalar Landau--Migdal interaction amplitude ${F}^{nn}(\bp_1,\bp_2)$ of two neutron quasi-particles with momenta $\bp_1,\bp_2$ is dominated by the contribution from an exchange of a static soft pion $\dl F^{nn}_{\pi}\propto D(\om=0,|\bp_1-\bp_2|)$ which has a strong momentum dependence \cite{Voskr2000}, see Eq.~(\ref{Dpi0c}). Several investigations have shown \cite{Voskr2000,VAKhod2004,PankrPRC2012,PankrJETPL2013} that this strong momentum dependence of the interaction amplitude ${F}^{nn}(\bp_1,\bp_2)$ can trigger topological phase transitions in neutron matter from the Landau state to states with more than one sheet of the Fermi surface. Thus, the study of pairing effects near the $\pi^0$ condensation point should include consideration of a possible non Fermi-liquid topology of the underlying ground state where the pairing correlations are switched off. A general discussion of pairing aspects in a Fermi system in a state with two sheets of the Fermi surface may be found in \cite{Clark2001MigAnn} where the method developed in \cite{KKC1996} was applied.

The investigation \cite{VAKhod2004} of nuclear pairing in dense neutron matter showed that the
spin-triplet $P$-wave neutron pairing is amplified by the soft pion exchange. In this article, we report on a possibility of spin-singlet $S$-pairing in the vicinity of the $\pi^0$ condensation point. We consider the pairing interaction induced by a soft pion and discuss specific features of the solution of the gap equation. This study is limited to the $^1S_0$ pairing in the Landau state. The extension of the study to the possible states with a non Fermi-liquid topology of the Fermi surface will be given elsewhere.

\section{Gap equation in the vicinity of the $\pi^0$-condensation critical point}

The general form of the many-body gap equation is as follows \cite{MigdalTKFS}: \beq \Delta = {\cal U} G G_s \Delta, \label{gap-eq}\eeq where ${\cal U}$ is the sum of interaction diagrams irreducible in the particle-particle channel. $G_s$ and $G$ are single-particle Green functions with and without pairing effects, respectively. The use of ${\cal U}=V_{NN}$, where $V_{NN}$ is the bare $NN$ potential, and a $G$-function with the free single-particle spectrum, corresponds to the BCS approximation. As was discussed in the Introduction, a lot of works was aimed to go beyond BCS. Firstly, they  included those devoted to the incorporation into ${\cal U}$ of the induced interactions, i.e. the sums of bubble diagrams \cite{SchulzePollsRamos2001,ShenLombardoSchuck2005,CaoLombardoSchuck2006,Dob}. Secondly,
contributions of the self-energy effects in Green functions were analyzed \cite{BaldoGrasso2000,LombardoSchuckZuo2001,BozekPRC2000,MutherDickhoff2005,Dob}. At last, the
retardation effects in the effective pairing interaction were also studied \cite{returd}. The soft pion induced interaction, $V_{\pi}\propto D$, which we discuss, appears due to a summation of bubble diagrams in the spin-isospin channel. As follows from Eq. (\ref{Dpi0c}), it has a singular form in the vicinity of the critical point. In this case, it is reasonable to separate this term from the interaction block ${\cal U}$, \beq {\cal U} = {\cal U}_{\rm reg} + V_{\pi}, \label{splitVpi}\eeq where the first term is the sum of all regular contributions. In the density region we consider, the term $V_{\pi}$ dominates over regular contributions.

In this study, we investigate the possibility of neutron pairing in the $^1S_0$ channel in the vicinity of the pion condensation critical point. For this aim,  we use a simple model omitting the term ${\cal U}_{\rm reg}$ in Eq. (\ref{splitVpi}). In addition, we use the static limit for the pion induced potential:\beq V_{\pi}(\bq) = \dfrac{{\tilde f}^2}{m_{\pi}^2}(\sg_1\bq)(\sg_2\bq) D(\om=0,\bq)\,.\label{Vpi_gen}\eeq Here the standard notation  \cite{PionBook} is used, and $\tilde f$ is the in-medium $\pi n$ coupling constant.

We use also the Green functions with the free single-particle spectrum, $\eps(p)=p^2/2m^*,\;\;m^*=m$, and renormalization factor $Z=1$. In the result, the initial gap equation (\ref{gap-eq}) is reduced to a BCS-like form. All corrections to this simplest approximation discussed above do  not change the singular form of the soft pion induced potential and result in a variation of its parameters only. In its turn, it may change  the density region where the pairing exists but does not cancel the phenomenon itself.  To estimate the role of the term ${\cal U}_{\rm reg}$ in Eq. (\ref{splitVpi}), we extended our model by including the free $NN$ potential, i.e. by solving the gap equation  with ${\cal U} = V_{NN} + V_{\pi}$. The results will be discussed in more detail below, but corrections to the gap values occurred to be rather small, $\sim 5\div 30$\%, depending on the density under consideration.

\section{Soft pion induced potential}

Projecting the potential (\ref{Vpi_gen}) onto the $nn$ spin-singlet state and taking into account Eq.~(\ref{Dpi0c}) for the propagator of the soft $\pi^0$, one arrives at the formula \beq V_{\pi}(\bq) = \dfrac{C_0\,g_\pi}{\left(q^2/q_c^2 - 1\right)^2 + \eta^2}\,,\label{Vpi_qspace}\eeq where $C_0=\nu_{\rm F}^{-1}=\pi^2/mp_{\rm F}$ is the inverse unrenormalized density of states, $g_{\pi}>0$ is an effective coupling constant and $\eta^2\propto(n_c-n)/n_c$ is a dimensionless measure of proximity of the system to the $\pi^0$ instability. In the following we will adopt the values \beq g_{\pi}=2.8\,,\quad\eta=\sqrt{4.6(n_c-n)/n_c}, \label{gkappa_model}\eeq obtained within a semi-microscopic model of the pion polarization operator considered in \cite{PankrPRC2012}. This model reproduces the realistic critical density $n_c\simeq0.2$ fm$^{-3}$ and the  model value $q_c=p_{\rm F}$ for the wave vector of the soft $\pi^0$ mode.

\begin{figure}[t]
\begin{center}
\includegraphics[width=1\linewidth,height=0.7\linewidth]{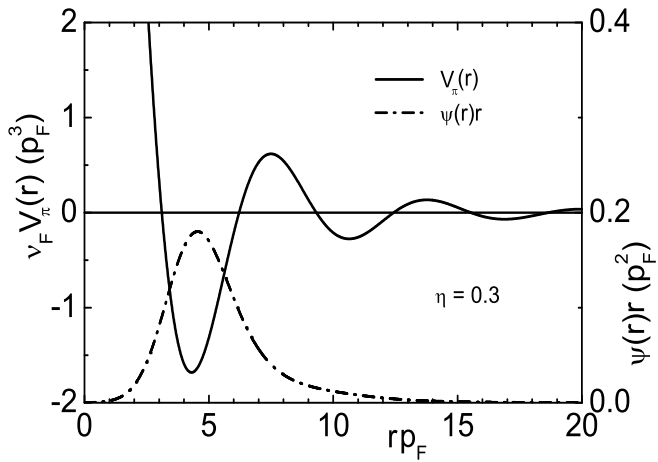}
\begin{minipage}[c]{0.9\linewidth}
\caption{The soft $\pi^0$ potential $V_{\pi}(r)$ in coordinate space in the spin-singlet state and the radial wave function $r\psi(r)$ of the lowest $S$-state at parameter $\eta=0.3$.} \label{fig_V_r}
\end{minipage}
\end{center}
\end{figure}

The spatial behavior of the Fourier transform $V_{\pi}(\br)$ of the potential (\ref{Vpi_qspace}) in coordinate space is shown in Fig.~\ref{fig_V_r}.
%
%
\begin{figure}[t]
\begin{center}
\includegraphics[width=0.95\linewidth,height=0.7\linewidth]{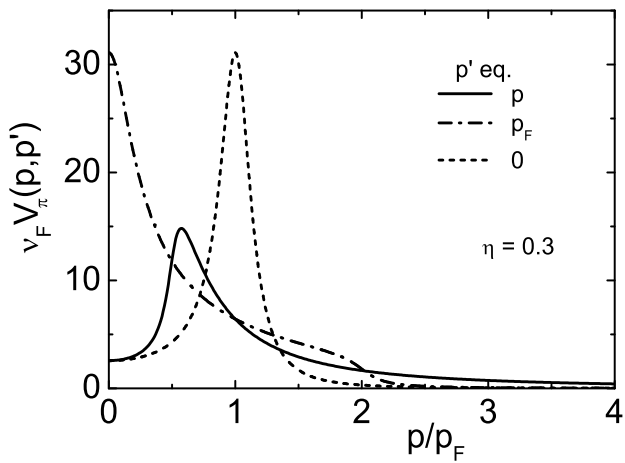}
\begin{minipage}[c]{0.9\linewidth}
\caption{The $S$-wave matrix elements $V_{\pi}(p,p')$ as a function of momentum: diagonal, $p'=p$; some off-diagonal, $p'=p_{\rm F}$ and $p'=0$.} \label{fig_Vqk}
\end{minipage}
\end{center}
\end{figure}
%
It has both regions of repulsion and attraction and resembles Fridel oscillations while its asymptotic form at distances $r>1/q_c$ is \beq V_{\pi}(\br) \simeq \dfrac{C_0g_{\pi}}{\eta}\dfrac{q_c^2}{4\pi r}\exp\left(-\dfrac{\eta q_c r}{2}\right)\left(\sin(q_c r)+\dfrac{\eta}{2}\cos(q_c r)\right)\,. \label{Vpi(r)}\eeq The potential supports bound states in the $S$ channel if the bare two-particle problem is considered. The first two $S$-levels appear when the parameter $\eta$ successively reaches the values $\eta_1\simeq0.55$ and $\eta_2\simeq0.24$. The radial wave function $r\psi_{\lambda}(r)$ of the bound $S$-state of the potential $V_{\pi}(\br)$ at $\eta=0.3$ is displayed in the same Fig.~\ref{fig_V_r}. The energy $\eps_{\lambda}$ of the lowest $S$-state eventually gets rather big values, of order the Fermi energy $p_{\rm F}^2/2m$, increasing in magnitude for decreasing $\eta$. The role of the bare bound $S$-state for the pairing problem that we consider is discussed in the following sections; the energy $\eps_\lambda$ and the average radius $\langle r_\lambda\rangle=\left(\int r^2|r\psi_\lambda|^2\,dr/\int|r\psi_\lambda|^2\,dr\right)^{1/2}$ of a bare bound pair for several values of $\eta$ are presented below.

The $S$-wave component of the potential $V_{\pi}(\bp-\bp')$ in momentum space is found by averaging of the expression (\ref{Vpi_qspace}) over the angle between momenta $\bp,\bp'$. The result is given in the explicit form: \begin{multline} V_{\pi}(p,p')=\dfrac{C_0g_{\pi}}{\eta}\dfrac{q_c^2}{4pp'}\left[
\arctan\dfrac{1}{\eta}\left(\left(\dfrac{p+p'}{q_c}\right)^2-1\right)\right.\\
\left.-\arctan\dfrac{1}{\eta}\left(\left(\dfrac{p-p'}{q_c}\right)^2-1\right)
\right]\,.\label{Vpiqk}\end{multline} Two specific properties of the matrix $V_{\pi}(p,p')$ are worth pointing out: {\it i}\,) all the matrix elements are positive; {\it ii}\,) the off-diagonal elements prevail over diagonal ones at small $\eta$. The first is obvious and the second follows from the comparison of the diagonal elements, $V_{\pi}(p,p)\propto\eta^{-1}$, with a representative off-diagonal one, $V_{\pi}(q_c,0)\propto\eta^{-2}$. Some of the matrix elements as a function of momentum are plotted in Fig.~\ref{fig_Vqk}. The dominant off-diagonality of the matrix $V_{\pi}(p,p')$ is the mathematical reason why a non-trivial solution of the gap equation appears in our case. We note that the similar situation holds for the Reid soft core $NN$ potential \cite{ReidSoftcore} well known in nuclear physics.

\section{Pairing correlations in the $^1S_0$ channel
in the vicinity of pion condensation point}

In vicinity of the critical point for the pion condensation, the parameter $\eta$ in the
denominators of Eqs. (\ref{Vpi(r)}) and (\ref{Vpiqk}) is small. It explains the dominance of this term over all regular components of the interaction block ${\cal U}$. In Fig.~\ref{fig_V_r}, the potential $V_{\pi}$ is displayed at $\eta=0.3$ which corresponds to $(n_c-n)/n_c\simeq 2\%$. The lowest minimum of this potential is much deeper, more than by an order of magnitude, than the one of the bare $NN$ potential such as the Argonne $v_{18}$ force \cite{Arg}. It explains why the simple model with ${\cal U}= V_{\pi}$ is reasonable in the density region under consideration. The accuracy of this approximation will be estimated below.

%
\begin{figure}[t]
\begin{center}
\includegraphics[width=0.9\linewidth,height=1.2\linewidth]{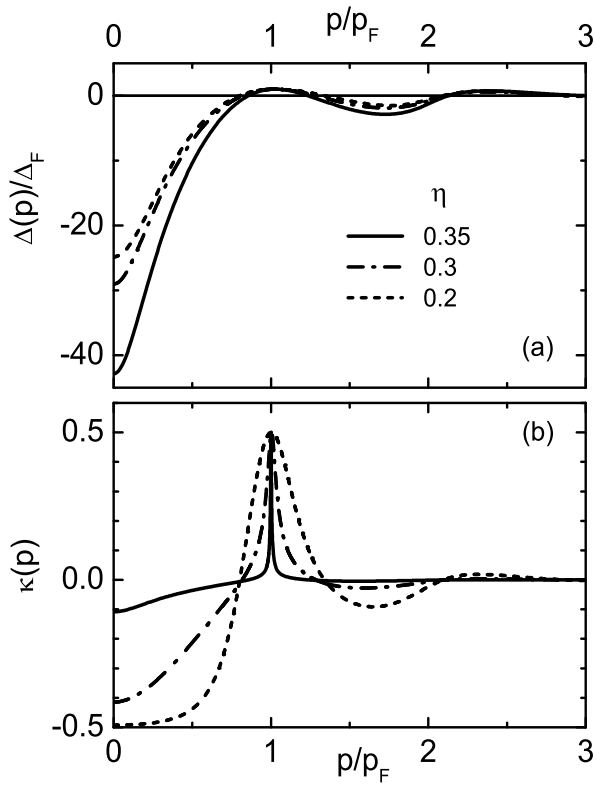}
\begin{minipage}[c]{0.9\linewidth}
\caption{(a) The gap function normalized by the value $\Dl_{\rm F}=\Dl(p_{\rm F})$ and (b) the anomalous density for several values of the parameter $\eta$.} \label{fig_D_chi_p}
\end{minipage}
\end{center}
\end{figure}
%

Thus, we solve the gap equation in the $^1S_0$ channel with the interaction (\ref{Vpiqk}), \beq \Dl(p)=-\int V_{\pi}(p,p')\,\dfrac{\Dl(p')}{2E(p')}\,\dfrac{p'^2dp'}{2\pi^2}\,,\label{gapeq}\eeq where $E(p)=\sqrt{(\eps(p)-\mu)^2+\Dl(p)^2}$ is the spectrum of Bogolubov quasiparticles. The superfluidity is regarded on the top of the Landau Fermi-liquid state with the quasiparticle momentum distribution $n_{\fl}(p)=\theta(p_{\rm F}-p)$ with one sheet of the Fermi surface at the Fermi momentum $p_{\rm F}=(3\pi^2n)^{1/3}$. As it was discussed in Section II, the spectrum $\eps(p)$ of initial quasiparticles of the nonsuperfluid state is fixed to be $\eps(p)=p^2/2m^*$ and the bare neutron mass $m^*=m$. The particle number conservation condition is used to find the chemical potential $\mu$: \beq\int{\cal N}(p)\,\dfrac{p^2dp}{\pi^2} = n\,,\label{norm}\eeq where ${\cal N}(p)$ is the momentum distribution of quasiparticles rearranged by pairing correlations: \beq {\cal N}(p)=\dfrac{1}{2}\left(1-\dfrac{\eps(p)-\mu}{E(p)}\right)\,.\label{Np}\eeq

The onset of superfluidity in neutron matter may be conveniently determined from the analysis of the linear equation \beq \left|2\eps(p)-p_{\rm F}^2/m\right|\kappa(p)=-\int V_{\pi}(p,p')\,\kappa(p')\,\dfrac{p'^2dp'}{2\pi^2}\,,\eeq that follows from Eq.~(\ref{gapeq}) in the limit of $\Dl\rightarrow0$. Here the anomalous density $\kappa(p)=\Dl(p)/2E(p)$ is introduced. The appearance of a non-trivial solution of this equation is the condition that the two-particle
scattering amplitude acquires a pole at the total momentum ${\bf P}=0$ and the energy $E=p_{\rm F}^2/m$ of a pair, as it follows from the in-medium Bethe--Salpeter equation. We found that the pairing instability in the $^1S_0$ channel occurs at $\eta_{\Delta}\simeq0.38$.

It is interesting to note that, since $\eta_{\Delta}<\eta_1\simeq0.55$, there is a range of values of the parameter $\eta$ where  a pair of particles interacting by means of the potential $V_{\pi}$ is bound being hypothetically placed in a vacuum while it is unbound in neutron matter. This situation is contrary to what one knows for the weak coupling attraction. The reasons why it takes place are the repulsion, $V_{\pi}(p_{\rm F},p_{\rm F})>0$, of the pairing interaction at the Fermi surface and the overlap of such pairs in neutron matter at densities we deal with. A more detailed discussion is given in the next section.

The normalized solution of the gap equation (\ref{gapeq}) is displayed in Fig.~\ref{fig_D_chi_p}(a). The gap function $\Dl(p)$ shows a strong momentum dependence. It gets maximum values {\it inside} the Fermi sphere and then changes its sign several times in order to satisfy the gap equation with the positive pairing interaction. The anomalous density $\kappa(p)$ corresponding to this gap function is plotted in Fig.~\ref{fig_D_chi_p}(b). One may see that, besides the usual sharp maximum attained at the Fermi surface $\left|\kappa(p_{\rm F})\right|=1/2$, the anomalous density also tends to this limit at the origin of the momentum axis as the parameter $\eta$ is decreased. This means that $\Dl(p)$ begins to prevail over the spectrum $\eps(p)-\mu$ in the inner region of the Fermi sphere.
%
\begin{table}[t]
\begin{minipage}[c]{0.9\linewidth}
\caption{The gap function at the points $p=0$ and $p=p_{\rm F}$, the variation of the chemical potential (both in units of $p_{\rm F}^2/2m$), and the correlation length. The energy (in units of $p_{\rm F}^2/2m$) and the average radius of a pair in the lowest bound $S$-state for the bare two-particle problem. All the quantities as functions of the parameter $\eta$. \label{Tab}}
\end{minipage}
\begin{tabular}{cccccccc}
\hline \hline \hspace*{2mm} $\eta$ \hspace*{2mm}&\hspace*{2mm} $\Dl_0$ \hspace*{2mm}&\hspace*{2mm}
$\Dl_{\rm F}$ \hspace*{2mm} & $\delta\mu \times 10$ &\hspace*{2mm} $\xi\,p_{\rm F}$
\hspace*{2mm} & \hspace*{2mm} $\eps_\lambda$ \hspace*{2mm} & $\langle r_\lambda\rangle\,p_{\rm F}$
\\
\hline
   0.35  &   -0.22  &    0.01  &    0.00  &  131.45  &  -0.25  &  5.17 \\
   0.32  &   -0.97  &    0.03  &    0.03  &   21.75  &  -0.37  &  4.99 \\
   0.30  &   -1.49  &    0.05  &    0.07  &   13.79  &  -0.47  &  4.91 \\
   0.27  &   -2.41  &    0.09  &    0.11  &    9.13  &  -0.66  &  4.82 \\
   0.25  &   -3.15  &    0.12  &    0.13  &    7.60  &  -0.82  &  4.78 \\
   0.22  &   -4.56  &    0.17  &    0.12  &    6.31  &  -1.13  &  4.72 \\
   0.20  &   -5.76  &    0.23  &    0.07  &    5.87  &  -1.41  &  4.69 \\
\hline \hline
\end{tabular}
\end{table}
%

We present details of the solution of the gap equation at different values of the parameter $\eta$ in Table~\ref{Tab}. The quantities $\Dl_0$ and $\Dl_{\rm F}$ are the values of the gap function $\Dl(p)$ at the points $p=0$ and $p=p_{\rm F}$, respectively. The variation of the chemical potential is defined as $\delta\mu=\mu-p_{\rm F}^2/2m$. The correlation length $\xi$ is introduced below. The above mentioned characteristics $\eps_\lambda$ and $\langle r_\lambda\rangle$ of the bare two-particle bound $S$-state are presented in the last two columns.

\begin{figure}[t]
\begin{center}
\includegraphics[width=0.9\linewidth,height=0.65\linewidth]{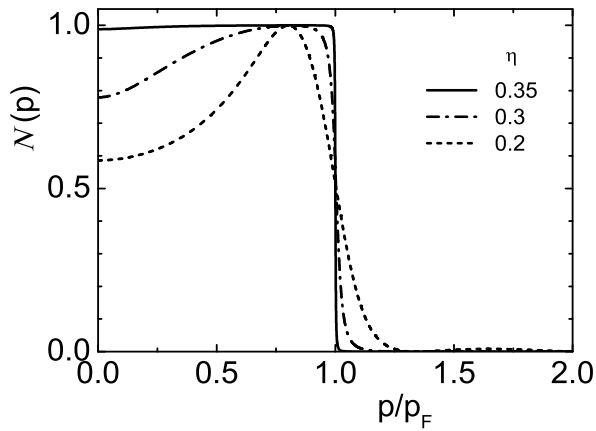}
\begin{minipage}[c]{0.9\linewidth}
\caption{Quasiparticle momentum distribution for several values of the parameter $\eta$.}
\label{fig_n}
\end{minipage}
\end{center}
\end{figure}

The strong momentum dependence of the gap function has specific influence on the
quasiparticle momentum distribution (\ref{Np}) which is plotted in Fig.~\ref{fig_n}.
%
%
This figure shows a dip in the occupation numbers ${\cal N}(p)$ at low momenta and a permanent
presence of a jump from 1 to zero in vicinity of the Fermi momentum for each value of the parameter $\eta$. At the same time, the chemical potential is almost unperturbed, $\mu\simeq p_{\rm F}^2/2m$, as one may see from Table~\ref{Tab}. We note that the found solution of the gap equation has no relation to the strong coupling limit despite of the presence of a bound state in the pairing potential. In the latter case one has $\mu<0$ and occupation numbers tend to be ${\cal N}(p)\ll1$ opening a possibility to Bose-Einstein condensation \cite{Nozieres_Schmitt-Rink}.
%
\begin{figure}[t]
\begin{center}
\includegraphics[width=0.9\linewidth,height=0.65\linewidth]{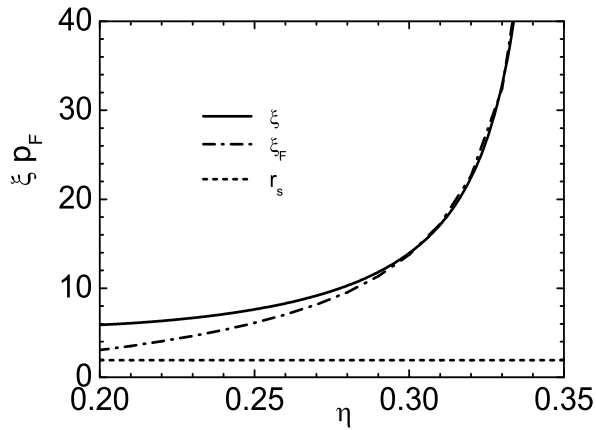}
\begin{minipage}[c]{0.9\linewidth}
\caption{The correlation length $\xi$ and its approximation $\xi_{\rm F}$ (both in units of $p_{\rm F}^{-1}$) as functions of the parameter $\eta$. The horizontal dotted line represents the average interparticle distance $r_s$ (same units).}\label{fig_xi}
\end{minipage}
\end{center}
\end{figure}
%

In order to clarify the character of pairing regime we calculated the correlation length \beq \xi = \sqrt{\dfrac{\int\left|\frac{\pd}{\pd
p}\kappa(p)\right|^2\,p^2dp}{\int\left|\kappa(p)\right|^2\,p^2dp}}\label{xi}\,.\eeq The dependence of this quantity on the parameter $\eta$ is given in Fig.~\ref{fig_xi}, as well as in Table~\ref{Tab}. It is clearly seen that the correlation length exceeds the average interparticle distance $r_s=(9\pi/4)^{1/3}/p_{\rm F}$ that implies a picture consistent with weak coupling. The figure also demonstrates a worsening of the usual estimate $\xi_{\rm F}=p_{\rm F}/m\sqrt{8}\Dl_{\rm F}$ of the correlation length with a decrease of the parameter $\eta$.  Thus, we deal here with a weak coupling regime with an unusually strong momentum dependence of the gap function inside the Fermi sphere.  The solution found by us may be related to a class of unconventional BSC solutions in the classification of the article \cite{Unconv} where an original point of view on pairing in neutron matter is presented.

Table~\ref{Tab} shows that the gap values we obtain are rather big in close vicinity to the pion
condensation critical point. In such a situation, the pairing changes the particle-hole propagator \cite{MigdalTKFS,Vaks,bigdelta}. Therefore the use of parameters of the potential $V_{\pi}$ found without this effect is questionable. However, this correction can not close the phenomenon we predict. Moreover, it does not change the critical density value for the pairing phase transition. Indeed, the latter can be found from the in-medium Bethe--Salpeter equation which does not contain the pairing gap.

To estimate validity of the approximation ${\cal U}=V_{\pi}$ in vicinity of the pion condensation critical point, we repeat the calculations for ${\cal U}=V_{\pi}+V_{NN}$ with the Argonne $v_{18}$ $NN$ force. The results are presented in Table~\ref{Tab2}. One may see that the effect of $V_{NN}$ does not exceed 30\% with the only exception of the $\eta=0.35$ point where the gap is close to zero.
\begin{table}[t]
\begin{minipage}[c]{0.9\linewidth}
\caption{The effect of the addition of the $NN$ potential $v_{18}$ to the soft pion potential
$V_{\pi}.$ The gap values are in units of $p_{\rm F}^2/2m$.\label{Tab2}}
\end{minipage}
\begin{tabular}{ccccc}
\hline \hline   $\eta$ \hspace*{2mm} & $\Dl_0[V_{\pi}]$ \hspace*{2mm} & $\Dl_{\rm F}[V_{\pi}]$
\hspace*{2mm}  & $\Dl_0[V_{\pi}{+}V_{NN}]$ \hspace*{2mm}   & $\Dl_{\rm F}[V_{\pi}{+}V_{NN}]$
\\
\hline
   0.35  &   -0.22   &   0.005  &  -0.44   &   0.013  \\
   0.32  &   -0.97   &   0.03   &  -1.15   &   0.04  \\
   0.30  &   -1.49   &   0.05   &  -1.66   &   0.06  \\
   0.27  &   -2.41   &   0.09   &  -2.56   &   0.10  \\
   0.25  &   -3.15   &   0.12   &  -3.29   &   0.13  \\
   0.22  &   -4.56   &   0.17   &  -4.69   &   0.19  \\
   0.20  &   -5.76   &   0.23   &  -5.88   &   0.25  \\

\hline \hline
\end{tabular}
\end{table}
%

\section{Discussion}

In this section we discuss in more detail several points concerning the solution of the gap equation with the interaction (\ref{Vpiqk}). The first point concerns the relation between the found weak coupling solution of the gap equation and the presence of a bound $S$-state in the pairing potential. In order to consider this problem it is useful to rewrite the gap equation in the following way \cite{Nozieres_Schmitt-Rink}:
\bea&2\left(\eps(p)-\mu\right)\kappa(p) = \hspace{43mm}&\\\notag &\hspace{0mm}-\left(1-2{\cal
N}(p)\right)\displaystyle\int V_{\pi}(p,p')\,\kappa(p')\,\dfrac{p'^2dp'}{2\pi^2}\,,&\\ &{\cal
N}(p)=\frac{1}{2}\left(1-\mathrm{sgn}(\eps(p)-\mu)\sqrt{(1-4|\kappa(p)|^2)}\right)\,.&\eea The
two-particle Schr$\ddot{\mathrm{o}}$dinger equation is then recovered for the {\it negative} chemical potential and $|\kappa(p)|\ll1$ when the approximate relation ${\cal N}(p)\simeq|\kappa(p)|^2\ll1$ holds. In this limit the normalization condition (\ref{norm}) takes the form $\int|\kappa(p)|^2\,p^2dp/\pi^2 = n$. Turning now to the pairing interaction (\ref{Vpi_qspace}) which has a characteristic radius $1/\eta p_{\rm F}$ one may conclude from the normalization condition that for a strongly  bound two-particle state $|\kappa(p)|^2\sim1/\eta^3>1$. Thus, the bound state solution cannot be properly normalized in order  not to violate  the constraint $|\kappa(p)|\ll1$. In other words the bound pairs of the Schr$\ddot{\mathrm{o}}$dinger equation overlap and the Pauli principle plays a significant role for the existence of the solution of the gap equation that we found.
%
\begin{figure}[t!]
\begin{center}
\includegraphics[width=0.9\linewidth,height=0.65\linewidth]{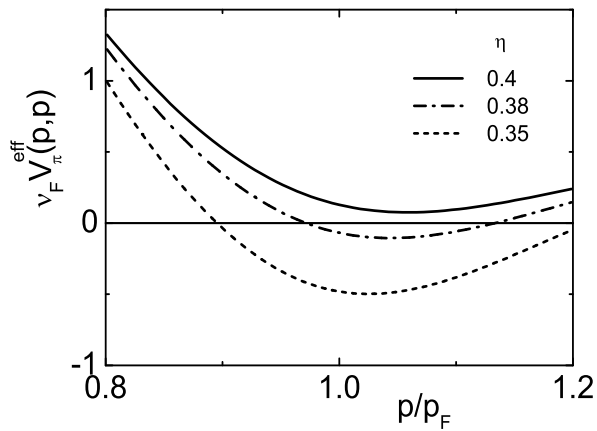}
\begin{minipage}[c]{0.9\linewidth}
\caption{Diagonal matrix elements of the effective pairing interaction found for the model space with $\dl p/p_{\rm F}=0.2$ for several values of the parameter $\eta$.}\label{fig_Veff}
\end{minipage}
\end{center}
\end{figure}
%

The second point is the specific role of the Fermi surface in the formation of a pairing gap in the case of the pairing interaction that is positive in momentum space. Despite the repulsion $V_{\pi}(p_{\rm F},p_{\rm F})>0$ of the pairing interaction on the Fermi surface it is that manifold in momentum space where the anomalous density $\kappa(p)$ appears from the outset. To understand this better, it is worth to consider the gap equation in the model space $S_0=\left(p_{\rm F}-\dl p,\,p_{\rm F}+\dl p\right)$ in terms of a renormalized pairing interaction \cite{MigdalTKFS}: \beq
\Dl(p)=-\int_{S_0}V_{\pi}^{\eff}(p,p')\,\dfrac{\Dl(p')}{2E(p')}\,
\dfrac{p'^2dp'}{2\pi^2}\,.\label{gapeqren}\eeq The renormalized interaction obeys the equation \beq V_{\pi}^{\eff}(p,p') = V_{\pi}(p,p') -
\int_{S'}\dfrac{V_{\pi}(p,q)V_{\pi}^{\eff}(q,p')}{2\left|\eps(q)-\mu\right|}\,
\dfrac{q^2dq}{2\pi^2}\label{Veff}\eeq in which the integration is carried out in the subspace $S'$ complementary to $S_0$. The model space $S_0$ has to be large enough for neglecting pairing effects $|\Dl(p)|\ll|\eps(p)-\mu|$ in the subspace $S'$. However, if one approaches the pairing instability from the superfluid side, $\Dl(p)\rightarrow0$, the model space may be chosen sufficiently small $\dl p/p_{\rm F}\ll1$. In this case the pairing gap in the model space can be obtained in the standard way: $\Dl_{\rm F}/\dl\eps = \exp(2/\nu_{\rm F}{\cal V}_{\rm F})$ where ${\cal V}_{\rm F}=V_{\pi}^{\eff}(p_{\rm F},p_{\rm F})$ and $\dl\eps=2p_{\rm F}\dl p/m$. The transition to a nonsuperfluid state is associated with the vanishing of the ${\it negative}$ effective pairing interaction ${\cal V}_{\rm F}\rightarrow0-$. Thus, one may see that the anomalous density $\kappa(p)$ concentrates in the model subspace of momentum space and shrinks with it to the Fermi surface as the superfluid phase transition is approached; see Fig.~\ref{fig_D_chi_p}b.  The effective interaction calculated from Eq.~(\ref{Veff}) for a narrow model space with $\dl p/p_{\rm F}=0.2$ is shown in Fig.~\ref{fig_Veff} for several values of the parameter $\eta$ near the critical one $\eta_\Delta\simeq0.38$ mentioned above.  We note that states away from the Fermi surface play an important role in the renormalization of the pairing interaction. Moreover, in our case all states inside the Fermi sphere must be included in the model space explicitly to obtain a correct solution for the strongly momentum dependent gap function at a distance beyond the superfluid transition.

\section{Conclusion}

We have examined a possibility of $^1S_0$ neutron pairing in dense neutron matter in the vicinity of the $\pi^0$ condensation point. The investigation was performed with the use of a simple BCS-like model with the pairing interaction induced by an exchange of the soft neutral pionic mode. Superfluidity was searched on the top of the Landau state with one sheet of the Fermi surface. It is shown that a non-trivial solution of the gap equation in the $^1S_0$ channel appears in a domain of $(n_c-n)/n_c\lesssim5\%$ near the critical density of the $\pi^0$ condensation. The gap function reveals a strong momentum dependence as the off-diagonal $S$-wave matrix elements of the pairing interaction prevail over diagonal ones while all the matrix elements are positive in momentum space. We have also discussed the weak coupling nature of the  superfluid phase that we found.

\section{Acknowledgments}

We are grateful to M.V. Zverev for useful discussions and remarks. Two of us, S.P. and E.S., thank INFN, Sezione di Catania, for hospitality during the stay in Catania when the major part of this work was done. This research was partially supported by Grant No. NSh-932.2014.2 of the Russian Ministry for Science and Education and by the RFBR Grants 12-02-00955-a, 13-02-00085-a, 13-02-12106-ofi\_m, 14-02-00107-a, 14-02-31353 mol\_a.

\end{document}